\documentclass[11pt, twocolumn]{article}
\usepackage{ElstonLabStyle}
\usepackage{cite}
\usepackage{float}
\usepackage{adjustbox}
\usepackage{natbib}

\usepackage{comment}
\usepackage{lipsum}
\usepackage{amsfonts}
\usepackage{amsmath}

\usepackage{amssymb}
\usepackage{chngcntr}
\usepackage[toc, page, title, titletoc]{appendix}
\usepackage{lineno}
\usepackage{microtype}

\usepackage{xr-hyper}

\usepackage[colorlinks=true, linkcolor=blue]{hyperref}
\usepackage{nameref}
\usepackage{cleveref}

\usepackage{xcolor}
\definecolor{LinkColor}{HTML}{197BD1} % Internal links (\ref, \eqref, section links)
\definecolor{CiteColor}{HTML}{8F5C25} % Bibliography citation links (\cite)
\definecolor{URLColor}{HTML}{197BD1}  % Web URLs (\url, \href)

\hypersetup{
    colorlinks=true,
    linkcolor=LinkColor,
    citecolor=CiteColor,
    urlcolor=URLColor
}

\title{Event timing precision under graded gene activation: an analytical approach}

\author[a]{Juan Sebasti\'an Hern\'andez}
\author[b]{Sayeh Rezaee}
\author[b]{C\'esar Nieto}
\author[c,$\dagger$]{Abhyudai Singh}

\affil[a]{Department of Physics, Universidad de los Andes, Bogot\'a, Colombia}
\affil[b]{Department of Electrical and Computer Engineering, University of Delaware, Newark, DE, USA}
\affil[c]{Department of Electrical and Computer Engineering, Biomedical Engineering, Mathematical Sciences, Center of Bioinformatics and Computational Biology, University of Delaware, Newark, DE, USA}
\affil[$\dagger$]{Correspondence: absingh@udel.edu}

\begin{document}

\begin{abstract}
The inherent stochasticity of gene expression causes cell-to-cell variability in the timing of events that are triggered when a protein reaches a threshold level. The statistics of this time, known as the first-passage time (FPT), have been characterized for single genes and for gene cascades with switch-like activation. However, real genes respond gradually to their activators, and how this graded activation shapes timing noise has so far been studied only numerically. Here, we characterize this effect analytically for a two-gene timing system. We model gene expression with the stochastic burst-dilution model and describe the activation of the downstream gene with a piecewise-linear function of the upstream protein level, defined by two parameters: a threshold and a slope. This approach reveals nontrivial optimal strategies, the combinations of threshold and slope that minimize timing noise at a fixed mean FPT. Our results, verified by stochastic simulations, show that timing is most precise when the activation saturates before or at the time of the event, and that low activation thresholds are best combined with graded responses, whereas higher thresholds favor responses that approach the switch-like limit. For identical genes, a graded response outperforms switch-like activation for every mean FPT we explored. The optimal threshold increases with the mean FPT, while the optimal slope varies non-monotonically. This work provides an analytically tractable framework for studying how graded activation affects timing precision, an approach that can be extended to longer activation cascades.

\end{abstract}

\maketitle

\section*{Introduction}

Multiple cellular processes are triggered when a protein accumulates to a critical level~\cite{grima_fpt}. For example, in the lytic pathway of bacteriophage~$\lambda$, the rate-limiting step for lysis of an infected \textit{E.~coli} cell is the accumulation of the holin protein to a threshold concentration in the inner membrane~\cite{white2011holin,cahill2024spatial,singh2014stochastic,mondal2024molecular,kannoly2025translation}. Similarly, in fission yeast, the nuclear concentration of the mitotic cyclin Cdc13 rises steadily during the cell cycle~\cite{vandal2026role}, and cells enter mitosis and divide when it exceeds a threshold~\cite{patterson2019noisy}. Because gene expression is stochastic~\cite{Thattai_vanOudenaarden_2001,Elowitz_Levine_Siggia_Swain_2002,Ozbudak_Thattai_Kurtser_Grossman_vanOudenaarden_2002,kaern2005stochasticity}, the time at which the protein first reaches its threshold, known as the first-passage time (FPT)~\cite{ghusinga2017first,iyerbiswas2016first}, varies from cell to cell~\cite{amir2007noise,ditalia2007effects}. This timing variability, quantified by the FPT noise, can affect cellular fitness~\cite{kannoly2022optimal,kannoly2025translation,nachman2007dissecting}, so natural selection can act on the timing noise of an event as well as on its mean time~\cite{kannoly2020optimum}. Identifying the parameter combinations that minimize FPT noise is therefore relevant for understanding how gene regulatory systems are tuned to time cellular events~\cite{ghusinga2017first, co2017stochastic}.

For timing systems built from a single gene, the FPT statistics are well characterized, and exact distributions are available~\cite{rijal2020protein,rijal2022exact}. With constitutive expression, the timing noise is lowest at an intermediate threshold~\cite{co2017stochastic}, a prediction confirmed in bacteriophage~$\lambda$, where the holin threshold of the wild-type phage is close to optimal~\cite{kannoly2020optimum}. When the protein is degraded or diluted, feedback on the gene's own production can lower the noise further~\cite{ghusinga2017first,rezaee2026event}. Other regulatory mechanisms, such as sequestration by decoys~\cite{Biswas_Dey_Singh_2024} and feedforward circuits~\cite{dey2021feedforward}, can also improve timing precision. In many systems, however, the gene that produces the triggering protein is not active from the start. Instead, its expression is activated by the protein of an upstream gene, which may in turn be activated by another gene. We refer to this arrangement as an activation cascade~\cite{rosenfeld2003response}. Regulation by an upstream activator can increase timing precision relative to the single-gene case~\cite{gupta2018temporal,gupta2020temporal}. In previous work, we derived expressions for the FPT noise of activation cascades of arbitrary length~\cite{hernandez2025cascade}. That analysis assumed switch-like activation, in which the downstream gene turns fully on as soon as the upstream protein crosses the activation threshold. Here, we address the more realistic case in which the production rate of the downstream gene increases gradually until full activation, a behavior known as a graded response~\cite{biggar2001cell}.

\begin{figure*}[t]
   \centering \includegraphics[width=0.9\textwidth]{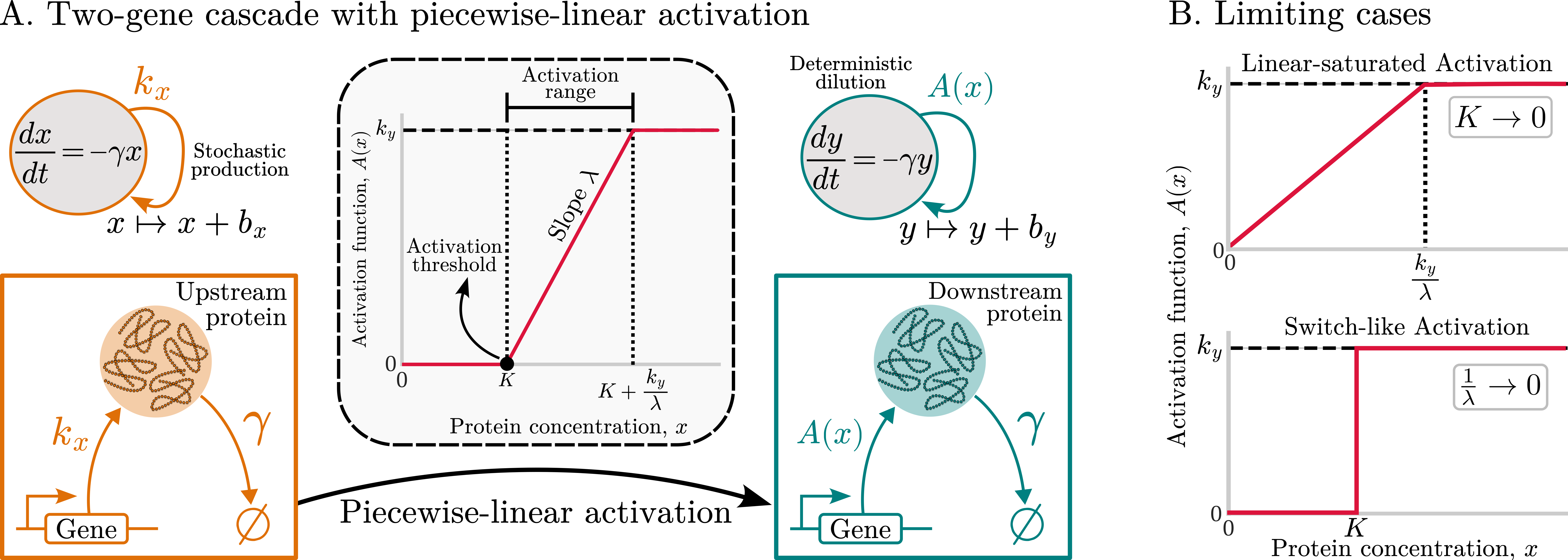}\caption{\small
    \textbf{Two-gene timing system with piecewise-linear activation.}
    (A) Schematic of the burst-dilution model. The upstream protein (concentration $x$) is produced constitutively in stochastic bursts of size $b_x$ that arrive at rate $k_x$, while the downstream protein (concentration $y$) is produced in bursts of size $b_y$ that arrive at rate $A(x)$. Both proteins are diluted at rate $\gamma$ [Eq.~\eqref{eq:dilution}]. The activation function $A(x)$ is characterized by two parameters. The activation threshold $K$ is the concentration of $x$ required to initiate production of the downstream gene. The slope $\lambda$ determines the activation range $k_y/\lambda$ over which the response increases linearly before saturating at rate $k_y$. Normalized by the steady-state concentration $\bar{x}$, the activation threshold and range become the dimensionless parameters in Eq.~\eqref{eq:dimless}.
    (B) Limiting cases. When $K \to 0$, the activation function becomes linear-saturated. When $\lambda \to \infty$ (equivalently, $1/\lambda \to 0$), the activation range collapses and the response becomes switch-like.
}
    \label{fig:fig1}
\end{figure*}

Graded responses are commonly modeled with Hill functions, which are typically sigmoidal~\cite{Bintu_Buchler_Garcia_Gerland_Hwa_Kondev_Kuhlman_Phillips_2005,Garcia_Phillips_2011}. With such a response, the production rate of the downstream gene follows the fluctuations of the upstream protein instead of switching on once. Part of the upstream noise is therefore transmitted to the downstream gene~\cite{pedraza2005noise,hooshangi2005ultrasensitivity}. Describing this transmission analytically is difficult because the nonlinearity of the Hill function prevents the moment equations of the downstream protein from closing~\cite{soltani2015moment,schnoerr2017approximation}. As a result, previous studies of graded activation have relied mostly on numerical methods. Gupta \textit{et al.}~\cite{gupta2018temporal} numerically computed the timing noise of a downstream gene regulated through a Hill function with fixed cooperativity, deriving closed-form approximations only in the switch-like limit. They later searched numerically for the optimal regulation function and found that it is not necessarily Hill-like~\cite{gupta2020temporal}. The downstream protein in these studies was only produced, with its degradation or dilution considered at most as a robustness check. Therefore, a closed-form description of how the shape of a graded activation function sets the timing noise of a diluted protein is still missing.

In this work, we obtain such a description by replacing the Hill function with a piecewise-linear activation function~\cite{quee2021ramp}. The downstream gene is off while the upstream protein is below the activation threshold. Above this threshold, its production rate rises linearly with a tunable slope until it saturates (Fig.~\ref{fig:fig1}A). This function retains the main characteristic of a graded response, since production increases over a finite range of upstream concentrations before saturating. Because it is linear in each of its three pieces, the moment equations of the burst-dilution model~\cite{singh2010stochastic} can be solved analytically, and we obtain approximations of the FPT noise for any activation threshold and slope (Secs.~\ref{sec:model} and~\ref{sec:fpt_noise}). These approximations agree with stochastic simulations across the parameter space we explored and capture how the optimal threshold and slope depend on the target mean FPT. For two identical genes, the optimal threshold increases with the mean FPT, whereas the optimal slope changes non-monotonically (Sec.~\ref{global_optimum}).

\begin{table*}[t]
\centering
\setlength{\belowcaptionskip}{6pt}
\caption{\small \textbf{Summary of the model.} Dynamics of the burst-dilution model and the main quantities used throughout, with the equations where they are defined.}
\label{tab:model}
\small
\setlength{\tabcolsep}{4pt}
\begin{tabular}{@{}llll@{}}
\hline
\multicolumn{4}{@{}l}{\textit{Dynamics}}\\
\multicolumn{2}{@{}l}{Upstream bursts: $x \to x+b_x$ at rate $k_x$} & \multicolumn{2}{l@{}}{Downstream bursts: $y \to y+b_y$ at rate $A(x)$}\\
\multicolumn{4}{@{}l}{Dilution between bursts: $dx/dt=-\gamma x$ and $dy/dt=-\gamma y$, Eq.~\eqref{eq:dilution}}\\
\hline
\multicolumn{2}{@{}l}{\textit{Gene expression}} & \multicolumn{2}{l@{}}{\textit{Activation}}\\
$k_x$ & Upstream burst rate & $A(x)$ & Activation function, Eq.~\eqref{activacion_piecewise}\\
$k_y$ & Maximum downstream burst rate & $K$ & Activation threshold\\
$\langle b_i\rangle$, $\langle b_i^2\rangle$ & Burst-size moments of gene $i=x,y$ & $\lambda$ & Activation slope\\
$\gamma$ & Dilution rate & $\alpha$ & Relative activation threshold, Eq.~\eqref{eq:dimless}\\
$\bar{x}$, $\bar{y}$ & Steady-state concentrations, Eq.~\eqref{eq:steady_states} & $\mu$ & Relative activation range, Eq.~\eqref{eq:dimless}\\
$\mathrm{CV}_x$, $\mathrm{CV}_y$ & Intrinsic noises, Eq.~\eqref{eq:intrinsic_noise} & & \\
\hline
\multicolumn{4}{@{}l}{\textit{Event timing}}\\
$Y$ & Downstream threshold & $T$ & FPT, Eq.~\eqref{eq:fpt_def}\\
$\gamma\langle T\rangle$ & Relative mean FPT & $\mathrm{CV}_T^2$ & FPT noise, Eq.~\eqref{eq:cvT}\\
\hline
\end{tabular}
\end{table*}

\section{Model formulation}\label{sec:model}

We consider the two-gene timing system shown in Fig.~\ref{fig:fig1}A. An upstream gene produces a protein with stochastic concentration $x(t)$ that activates the expression of a downstream gene, whose protein has concentration $y(t)$. Both concentrations start from zero. The timing of the system is characterized by the FPT, the time at which $y$ first reaches the downstream threshold $Y$ that triggers the cellular event.

To describe gene expression, we use the \textit{burst-dilution} model~\cite{singh2010stochastic}, a piecewise-deterministic Markov process. In this model, transcription and translation are simplified as abrupt protein synthesis events (bursts) in which a random amount of protein is produced instantaneously~\cite{Cai_Friedman_Xie_2006,Yu_Xiao_Ren_Lao_Xie_2006,friedman2006linking}. The burst sizes $b_i$ ($i=x,y$) are i.i.d.\ positive increments in protein concentration, drawn from distributions with moments $\langle b_i \rangle$ and $\langle b_i^2 \rangle$, where $\langle \cdot\rangle$ denotes the mean value. For the upstream gene, bursts arrive as a Poisson process with rate $k_x$. For the downstream gene, they arrive at a rate given by the activation function $A(x)$, which couples the upstream concentration to downstream expression. Between bursts, both concentrations decay deterministically through dilution by cell growth, at a common rate $\gamma$
\begin{equation}\label{eq:dilution}
\frac{dx}{dt} = -\gamma x, \qquad \frac{dy}{dt} = -\gamma y.
\end{equation}
The model and its parameters are summarized in Table~\ref{tab:model}.

\subsection{Activation function}

We define the activation function $A(x)$ as the piecewise-linear function
\begin{equation}\label{activacion_piecewise}
A(x):=\left\{
\begin{array}{ll}
0\; &\text{if } \;\;x\leq K;\\
\lambda (x-K) \; & \text{if } \;\;K<x<K+\frac{k_y}{\lambda};\\
k_y\; \; &  \text{if } \;\; x\geq K+\frac{k_y}{\lambda}.
\end{array}
\right.
\end{equation}
Thus, the downstream gene is inactive for $x \leq K$. Once $x$ exceeds the activation threshold $K > 0$, the production rate increases linearly with slope $\lambda$ until it saturates at the maximum rate $k_y$ (Fig.~\ref{fig:fig1}A). In the limit $K \to 0$, the function reduces to a linear-saturated form, whereas in the limit $\lambda \to \infty$ it becomes switch-like (Fig.~\ref{fig:fig1}B).

\subsection{Moment dynamics and dimensionless parameters}

The time evolution of the moments follows from Dynkin's formula~\cite{singh2010stochastic}, which gives the rate of change of the mean of any function $f(x,y)$ of the protein concentrations. Taking $f$ equal to $x^m$, $y^n$, and $x^m y^n$ ($m,n\in\mathbb{N}$) for the model in Table~\ref{tab:model} yields the moment dynamics
\begin{subequations}\label{eq:dynkin}
\begin{align}
\frac{d\langle x^m \rangle}{dt} &= k_x \langle (x + b_x)^m - x^m \rangle - \gamma m\langle  x^{m} \rangle, \label{eq:dynkin_x}\\[1ex]
\frac{d\langle y^n \rangle}{dt} &= \langle A(x) [(y + b_y)^n - y^n] \rangle - \gamma n\langle y^n \rangle, \label{eq:dynkin_y}\\
\frac{d\langle x^my^n \rangle}{dt} &= \langle k_x y^n[(x + b_x)^m - x^m] \rangle \nonumber\\
&\quad+\langle A(x) x^m[(y + b_y)^n - y^n] \rangle \nonumber\\
&\quad- \gamma (m+n)\langle x^m y^n\rangle. \label{eq:dynkin_xy}
\end{align}
\end{subequations}
Solving Eq.~\eqref{eq:dynkin_x} for $m=1$ and $m=2$ with $x(0)=0$ gives the mean $\langle x \rangle$ and the variance $\sigma^2_x=\langle x^2\rangle-\langle x\rangle^2$ of the upstream protein
\begin{equation}\label{eq:x_moments}
    \langle x \rangle = \bar{x}\left(1 - e^{-\gamma t}\right), \qquad
    \sigma^2_x = \frac{\langle b_x^2 \rangle}{2\langle b_x \rangle}\bar{x}\left(1-e^{-2\gamma t}\right),
\end{equation}
where $\bar{x}$ is the steady-state concentration. When the downstream gene is fully activated, its bursts arrive at the constant rate $k_y$, and its concentration approaches the steady state $\bar{y}$. These steady-state concentrations are
\begin{equation}\label{eq:steady_states}
\bar{x} := \frac{k_x \langle b_x \rangle}{\gamma}, \qquad \bar{y} := \frac{k_y \langle b_y \rangle}{\gamma}.
\end{equation}
We characterize each gene by its \textit{intrinsic noise}, defined as the steady-state coefficient of variation (standard deviation divided by the mean) of its protein concentration, with the downstream gene taken as fully activated. From Eq.~\eqref{eq:x_moments} and its analog for the downstream gene, the squared intrinsic noises are
\begin{equation}\label{eq:intrinsic_noise}
\begin{aligned}
\mathrm{CV}_x^2 &:= \lim_{t\to\infty}\frac{\sigma_x^2}{\langle x\rangle^2} = \frac{\langle b_x^2\rangle}{2\langle b_x\rangle \bar{x}},\\
\mathrm{CV}_y^2 &:= \lim_{t\to\infty}\frac{\sigma_y^2}{\langle y\rangle^2} = \frac{\langle b_y^2\rangle}{2\langle b_y\rangle \bar{y}}.
\end{aligned}
\end{equation}
We normalize the activation threshold and the activation range by $\bar{x}$, which gives the dimensionless parameters
\begin{equation}\label{eq:dimless}
\alpha := \frac{K}{\bar{x}}, \qquad \mu := \frac{k_y}{\lambda\, \bar{x}}.
\end{equation}
Here, $\alpha$ is the \textit{relative activation threshold} and $\mu$ is the \textit{relative activation range}, the width of the linear response region as a fraction of $\bar{x}$ (Fig.~\ref{fig:fig1}A). On average, the upstream concentration crosses the activation threshold when $\langle x \rangle / \bar{x}$ reaches $\alpha$, and the activation function saturates when $\langle x \rangle / \bar{x}$ reaches $\alpha + \mu$. Whether $y$ crosses the downstream threshold before or after the activation saturates defines two dynamical regimes with different noise properties.

\subsection{Estimation of FPT statistics}\label{sec:fptstats}

Formally, the FPT is the random variable
\begin{equation}\label{eq:fpt_def}
    T := \inf\{t \,:\, y(t) \geq Y \mid x(0) = 0,\, y(0) = 0\}.
\end{equation}
As with the parameters in Eq.~\eqref{eq:dimless}, we use the dimensionless \textit{relative mean FPT}, $\gamma \langle T \rangle$, which measures the mean event time in units of the dilution timescale $1/\gamma$.

Setting $\langle x \rangle/\bar{x} = \alpha$ and $\langle x \rangle/\bar{x} = \alpha+\mu$ in Eq.~\eqref{eq:x_moments} gives the times at which the mean upstream concentration reaches the activation threshold (activation time $t_a$) and the saturation level (saturation time $t_s$)
\begin{equation}\label{eq:times}
    t_a := \frac{1}{\gamma}\ln\!\left(\frac{1}{1-\alpha}\right), \quad t_s := \frac{1}{\gamma}\ln\!\left(\frac{1}{1-\alpha-\mu}\right).
\end{equation}
Because the downstream gene starts producing only after the activation time, a target mean FPT requires $\langle T\rangle > t_a$, which sets an upper bound on $\alpha$
\begin{equation}\label{eq:alpha_max}
    \alpha_{\max} := 1 - e^{-\gamma\langle T\rangle}.
\end{equation}
If $t_a < \langle T\rangle < t_s$, the event occurs while the activation is still rising linearly (partial activation regime). If instead $\langle T\rangle > t_s$, the event occurs after the activation has saturated (full activation regime). For a given $\alpha$, full activation requires
\begin{equation}\label{eq:regime_boundary}
    \mu < 1-\alpha-e^{-\gamma\langle T\rangle},
\end{equation}
and we call the case of equality the regime boundary.

To quantify the FPT noise, we use the squared coefficient of variation
\begin{equation}\label{eq:cvT}
    \mathrm{CV}_T^2 := \frac{\sigma_T^2}{\langle T \rangle^2},
\end{equation}
where $\sigma_T^2$ is the variance of $T$. In the small-noise limit, the FPT noise can be approximated as~\cite{co2017stochastic}
\begin{equation}\label{relating_protein_FPT}
\mathrm{CV}_T^2 \approx \lim_{t\rightarrow\langle T\rangle}\frac{\sigma^2_y}{\langle T\rangle^2} \left( \frac{d \langle y \rangle}{dt} \right)^{-2}.
\end{equation}
This approximation reduces the calculation of the FPT noise to evaluating the variance $\sigma^2_y$ of $y$ and the rate of change of its mean $\langle y\rangle$ at $t = \langle T \rangle$. In turn, the mean FPT is approximated by the time at which $\langle y\rangle$ reaches the downstream threshold
\begin{equation}\label{eq:mean_fpt}
    \langle y\rangle\big|_{t=\langle T\rangle} \approx Y.
\end{equation}
\begin{figure}[H]
    \centering  \includegraphics[width=1.0\columnwidth]{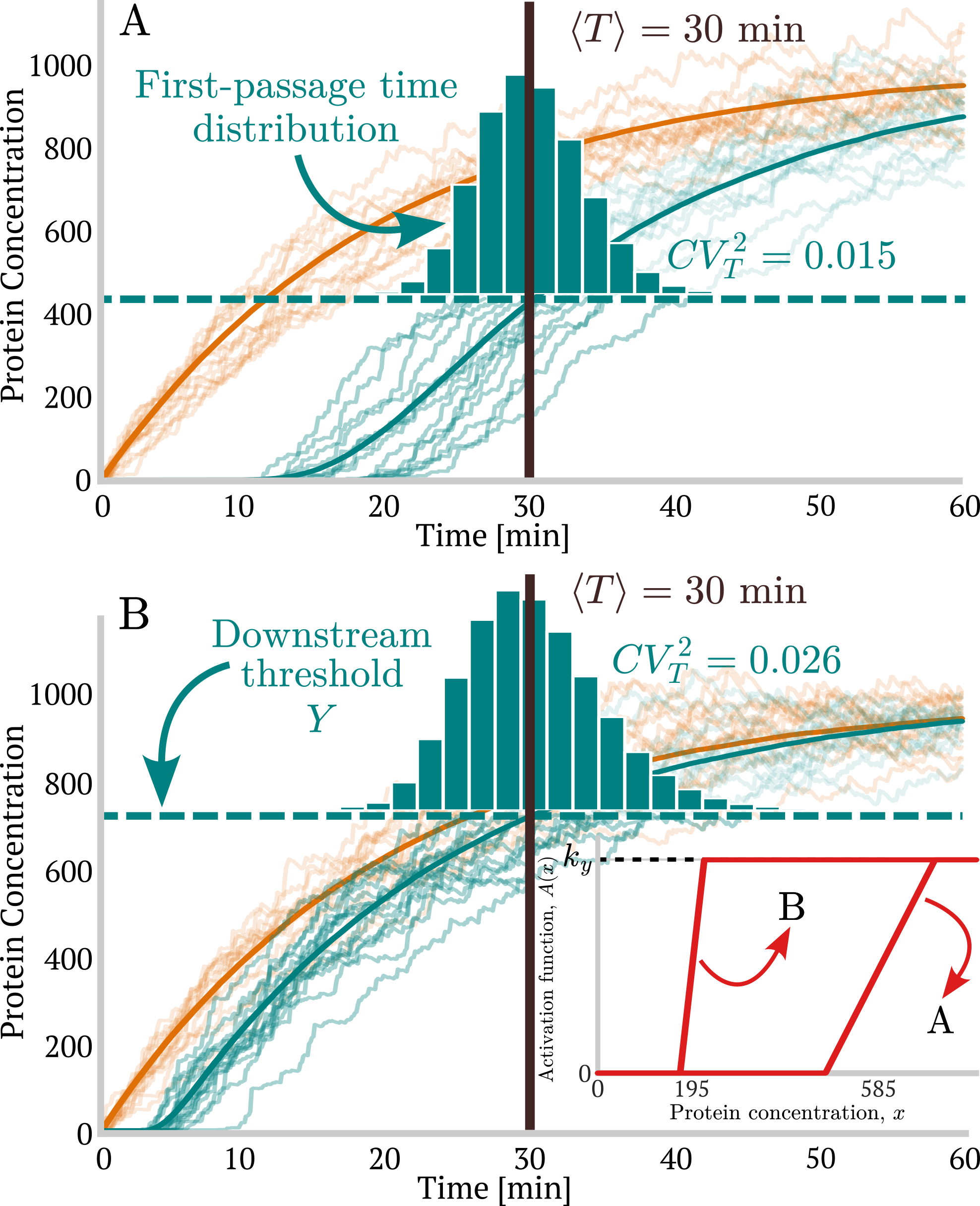}
    \caption{\small
    \textbf{Activation functions with the same mean FPT produce different timing noise.}
    Stochastic trajectories (semi-transparent lines) and mean protein concentrations (solid lines) of the upstream (orange) and downstream (teal) genes for two activation functions (panels A and B). The two activation functions, shown in the inset, share the same maximum rate $k_y$ but differ in their threshold $K$ and slope $\lambda$. In both cases, the downstream threshold $Y$ is adjusted so that the mean FPT is the same ($\langle T\rangle = 30\,\text{min}$, vertical line). However, the two FPT distributions (teal histograms) have different widths, with $\mathrm{CV}_T^2=0.015$ in A and $0.026$ in B. Thus, for a fixed $\langle T\rangle$, the shape of the activation function controls timing precision. \textbf{Parameters:} $\gamma = 0.05\,\text{min}^{-1}$, $k_y=10\,\text{min}^{-1}$, $\langle b_x\rangle =\langle b_y\rangle = 5$, $\langle b_x^2\rangle =\langle b_y^2\rangle = 50$, $\bar{x} = 1000$, $\gamma \langle T\rangle = 1.5$. Histograms are based on $10^5$ Monte Carlo samples.}
\label{fig:comparison}
\end{figure}

\section{FPT noise for piecewise-linear activation}\label{sec:fpt_noise}

We derive analytical expressions for the FPT noise $\mathrm{CV}_T^2$ as a function of $\alpha$, $\mu$, and $\gamma\langle T\rangle$ in each of the two activation regimes. The full derivation of the moments of $y$ is given in Appendix~\ref{app:moments}.

\subsection{Regime of partial activation ($t_a<\langle T\rangle <t_s$)}\label{sec:partial}

In this regime, the downstream gene is driven only by the linear piece of the activation function, $A(x) = \lambda(x - K)$. Solving the moment equations~\eqref{eq:dynkin} up to second order for $t_a<t<t_s$ gives the mean and variance of $y$
\begin{subequations}
\begin{align}
\label{eq:mean_y_lin}
\langle y \rangle =& \frac{\lambda\langle b_y\rangle(1-\alpha)\bar{x}}{\gamma}\left[1 - (1 + \gamma \tau)e^{-\gamma \tau}\right],\\
\label{eq:var_y_lin}
\sigma^2_y =\;& B \left[ \frac{\gamma(1-\alpha)\langle b_y^2\rangle}{\lambda \bar{x} \langle b_y \rangle^2} \left(1 - e^{-\gamma\tau}\right)^2 \right. \nonumber\\
&\left. + \mathrm{CV}_x^2\left(1 - \left(1 + 2\gamma\tau + 2\gamma^2(1-\alpha)^2\tau^2\right)e^{-2\gamma\tau} \right)\right],
\end{align}
\end{subequations}
where $\tau := t - t_a$ and $B :=\lambda^2 \langle b_y \rangle^2 \bar{x}^2/(2\gamma^2)$ is a common prefactor. Applying Eq.~\eqref{relating_protein_FPT}, we obtain the FPT noise in this regime
{\small
\begin{align}\label{eq:cv2_unsat}
    \mathrm{CV}^2_T \approx \frac{1}{(\gamma\langle T \rangle)^2}\bigg[&\mathrm{CV}_x^2\left(\frac{(1-\alpha)^2 e^{2\gamma \langle T\rangle} - 1 - 2s}{2(1-\alpha)^2 s^2} - 1\right) \nonumber\\
    &+ \mathrm{CV}_y^2\frac{\mu}{1-\alpha}\cdot\frac{\left((1-\alpha)e^{\gamma \langle T\rangle}-1\right)^2}{s^2}\bigg],
\end{align}
}
where $s := \gamma (\langle T \rangle-t_a)$ is the time elapsed since activation in units of $1/\gamma$. Because Eq.~\eqref{eq:cv2_unsat} depends linearly on $\mu$ with a strictly positive coefficient, the FPT noise in this regime increases monotonically with $\mu$ at fixed $\gamma\langle T\rangle$ and~$\alpha$.

\subsection{Regime of full activation ($\langle T\rangle>t_s$)}

In this regime, once the activation saturates, $y$ is produced constitutively at rate $k_y$, and its moments decouple from $x$. Solving Eq.~\eqref{eq:dynkin} for $t>t_s$, with the initial value $\langle y(t_s)\rangle$ given by Eq.~\eqref{eq:mean_y_lin}, yields the mean of $y$
\begin{equation}\label{eq:mean_y_sat}
\langle y \rangle = \bar{y}\left(1 - \frac{\bar{f}}{\mu}\,e^{-\gamma(t-t_s)}\right),
\end{equation}
where
\begin{equation}\label{eq:fbar}
\bar{f} := (1-\alpha-\mu)\ln\!\left(\frac{1-\alpha}{1-\alpha-\mu}\right).
\end{equation}
Similarly, the variance of $y$ is
\begin{equation}\label{eq:var_y_sat}
\sigma^2_y = \frac{k_y\langle b_y^2\rangle}{2\gamma}\left(1 - e^{-2\gamma(t-t_s)}\right) + \sigma^2_y(t_s)e^{-2\gamma(t-t_s)},
\end{equation}
where $\sigma^2_y(t_s)$ is Eq.~\eqref{eq:var_y_lin} evaluated at $t=t_s$. The corresponding FPT noise follows from Eq.~\eqref{relating_protein_FPT}
{\footnotesize
\begin{align}\label{eq:cv2_sat}
    \mathrm{CV}^2_T \approx& \frac{1}{(\gamma\langle T \rangle)^2}\bigg[\mathrm{CV}_x^2 \left( \frac{\mu}{2(1-\alpha)\bar{f}^2}\!\left(2-\frac{\mu}{1-\alpha}\right)-\frac{1-\alpha-\mu}{(1-\alpha)^2\bar{f}}-1 \right)
    \nonumber
    \\
    &+\mathrm{CV}_y^2 \left( \frac{\mu}{\bar f}\right)^2\!\left( (1-\alpha-\mu)^2 e^{2\gamma \langle T \rangle}+\frac{\mu}{1-\alpha}-1 \right) \bigg].
\end{align}
}
In both Eqs.~\eqref{eq:cv2_unsat} and~\eqref{eq:cv2_sat}, the term proportional to $\mathrm{CV}_x^2$ is the \textit{upstream contribution} to the FPT noise, which arises from fluctuations in the upstream protein. The term proportional to $\mathrm{CV}_y^2$ is the \textit{downstream contribution}, which arises from fluctuations in the expression of the downstream gene itself.

\begin{figure}[H]
    \centering  \includegraphics[width=1.0\columnwidth]{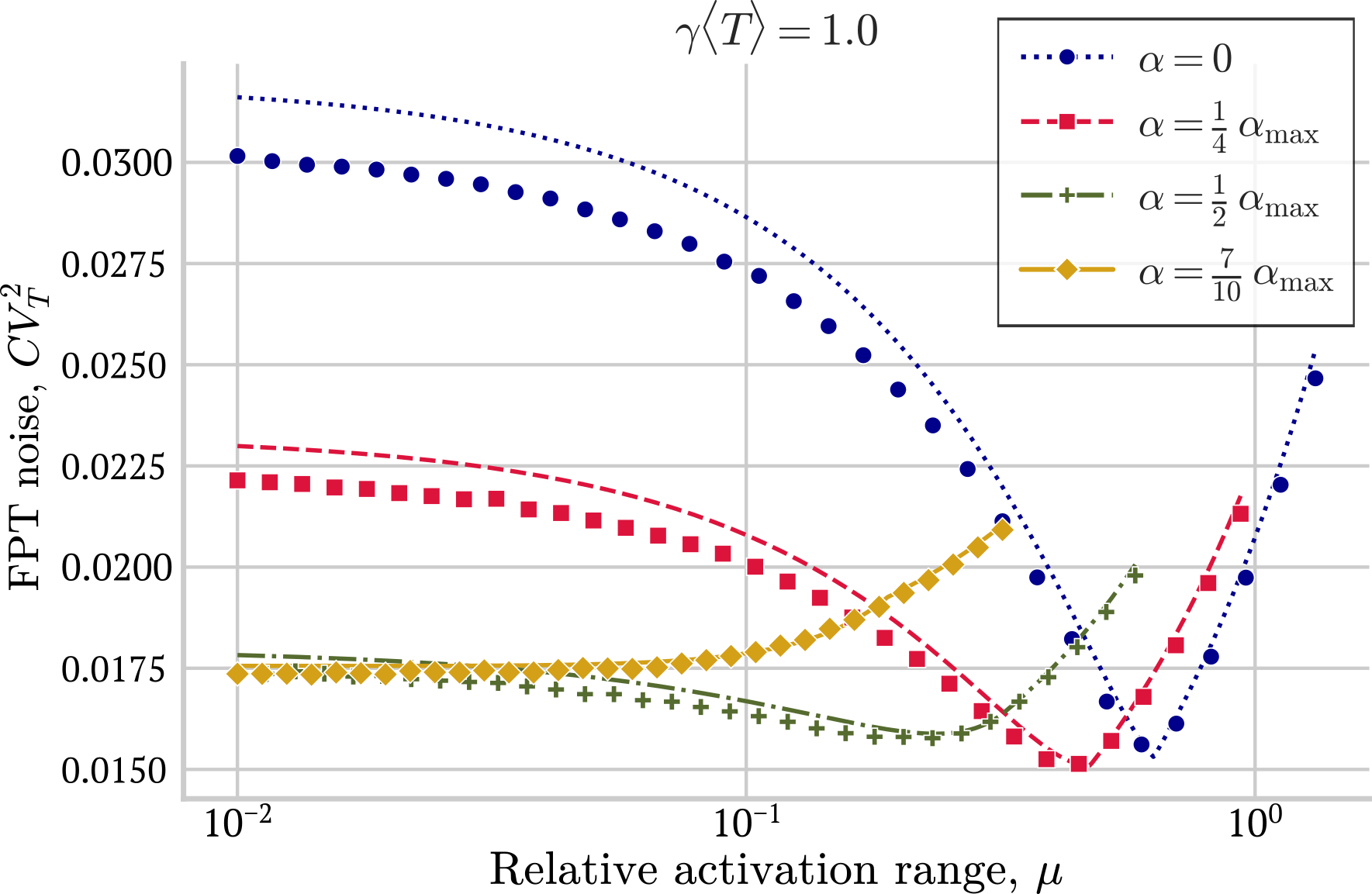}
   \caption{\small
       \textbf{The analytical expressions capture how the FPT noise changes with the activation range.} FPT noise $\mathrm{CV}_T^2$ as a function of $\mu$ for different values of $\alpha$ at fixed $\gamma\langle T\rangle = 1.0$. Lines show the analytical expressions~\eqref{eq:cv2_unsat} and~\eqref{eq:cv2_sat}, and markers show Monte Carlo simulations ($10^6$ samples). For each $\alpha$, $\mu$ ranges from values at which the system remains in partial activation (large $\mu$) to a nearly switch-like response (small $\mu$). The values of $\alpha$ are given as fractions of $\alpha_{\max}\approx 0.632$ [Eq.~\eqref{eq:alpha_max}], the largest threshold compatible with the target $\gamma\langle T\rangle$. \textbf{Parameters:} $\gamma = 0.05\,\text{min}^{-1}$, $k_y=10\,\text{min}^{-1}$, $\langle b_x\rangle =\langle b_y\rangle = 5$, $\langle b_x^2\rangle =\langle b_y^2\rangle = 50$, $\bar{x} = 1000$.}\label{fig:analytical_simulations}
\end{figure}

\subsection{Optimization over $\mu$ in both activation regimes}\label{sec:opt_mu}

With these expressions, we find the relative activation range $\mu^*$ that minimizes the FPT noise and analyze how it depends on the relative activation threshold $\alpha$. For simplicity, we focus on identical genes, which share the same burst statistics ($b_x \overset{d}{=} b_y$) and steady-state levels ($\bar{x} = \bar{y}$), so that $\mathrm{CV}_x^2 = \mathrm{CV}_y^2$. With $\bar{y}$, $\langle b_y\rangle$, and $\gamma$ fixed, $k_y$ is fully determined by Eq.~\eqref{eq:steady_states}, so varying $\mu$ is equivalent to varying the activation slope $\lambda$. For each pair $(\alpha, \mu)$, we adjust the downstream threshold $Y$ to keep the mean FPT $\langle T \rangle$ fixed (Fig.~\ref{fig:comparison}). This constraint ensures a fair comparison between activation functions, since all of them then trigger the event at the same mean time.

Figure~\ref{fig:analytical_simulations} shows the FPT noise as a function of $\mu$ for several values of $\alpha$. Equations~\eqref{eq:cv2_unsat} and~\eqref{eq:cv2_sat} are continuous at the regime boundary, but their derivatives with respect to $\mu$ are not, which produces a kink in each curve where the system switches from one regime to the other. Because the FPT noise increases monotonically with $\mu$ in the partial activation regime (Sec.~\ref{sec:partial}), $\mu^*$ always lies in the full activation regime or on its boundary. Within this regime, we distinguish three cases, depending on $\alpha$:

\begin{itemize}
    \item \textit{Low $\alpha$:} Activation starts early, long before the upstream protein reaches its steady state, and the upstream contribution is small. The optimum therefore approaches the widest activation range that still saturates before the event, given by the upper bound in Eq.~\eqref{eq:regime_boundary}.

    \item \textit{High $\alpha$:} Activation starts late, when the fluctuations in $x$ are large, and a switch-like response ($\mu^* \to 0$) minimizes the exposure to them. In this limit, upstream fluctuations only shift the time at which the downstream gene turns on, without modulating its production rate.

    \item \textit{Intermediate $\alpha$:} The optimal range $\mu^*$ balances the upstream and downstream contributions to the FPT noise and lies between these two extremes.
\end{itemize}

These three cases arise because the activation range redistributes the FPT noise between its two contributions. With a wider activation range, the production of $y$ accelerates gradually, so its mean is steeper when it crosses the threshold and the downstream contribution decreases. At the same time, a wider range lets upstream fluctuations modulate the production rate of $y$ for longer, which increases the upstream contribution. The higher the activation threshold, the larger the upstream contribution already is, so the balance shifts toward narrower activation ranges and $\mu^*$ decreases toward the switch-like limit. The activation threshold and range therefore cannot be chosen independently: a low threshold is best paired with a graded response, and a high threshold with a switch-like one.

Compared with the simulations in Fig.~\ref{fig:analytical_simulations}, the analytical expressions overestimate the noise at low values of $\alpha$ and $\mu$. In this region, $Y$ is a large fraction of $\bar{y}$, so $d\langle y\rangle/dt$ is small at the crossing. Because the small-noise approximation~\eqref{relating_protein_FPT} divides by the square of this rate of change, it predicts a large FPT noise. In the simulations, however, upward fluctuations of $y$ carry many cells above the threshold well before the mean reaches it, so the FPT noise in the simulations is smaller than predicted. Nevertheless, the expressions capture the location of the minimum and the qualitative shape of the noise curves.

\subsection{Optimal strategy in the limiting cases}

We now determine the optimal strategy, that is, the combination of activation parameters that minimizes the FPT noise, in the two limits of the activation function (Fig.~\ref{fig:fig1}B). In the limit $\lambda \to \infty$ ($\mu \to 0$), the activation becomes switch-like, the timing splits into two independent stages, and Eq.~\eqref{eq:cv2_sat} reduces to the two-gene result of our previous work~\cite{hernandez2025cascade}. For two identical genes, $\mathrm{CV}_T^2$ is then convex in $\alpha$, and the optimal threshold for a given $\gamma\langle T\rangle$ is
\begin{equation}\label{eq:alpha_opt_switch}
    \lim_{\mu \to 0} \alpha^* = 1 - e^{-\frac{\gamma\langle T\rangle}{2}}.
\end{equation}
With this threshold, the upstream protein reaches the activation threshold at $t_a = \langle T\rangle/2$, so the mean FPT is split evenly between the two genes.
% For non-identical genes, the optimal strategy depends on the ratio $\mathrm{CV}_y/\mathrm{CV}_x$, determining whether one or both genes should be used~\cite{hernandez2025cascade}. Whether these regimes persist under graded activation is a central question for future work.

In the opposite limit $K \to 0$ ($\alpha \to 0$), the downstream gene starts producing as soon as the upstream protein appears. For the values of $\gamma\langle T\rangle$ explored here, the optimal activation range then lies at the regime boundary (Fig.~\ref{fig:analytical_simulations})
\begin{equation}\label{eq:mu_opt_linear}
    \lim_{\alpha \to 0} \mu^* \approx 1 - e^{-\gamma\langle T\rangle}.
\end{equation}
With this activation range, the production rate of the downstream gene rises throughout the waiting time and saturates exactly when the event occurs. This contrasts with the switch-like limit, where the mean FPT is split between an upstream and a downstream stage. In the linear-saturated limit, both genes are active from the start, and the optimal strategy spends no time at full activation.

\begin{figure*}[t]
   \centering \includegraphics[width=1.0\textwidth]{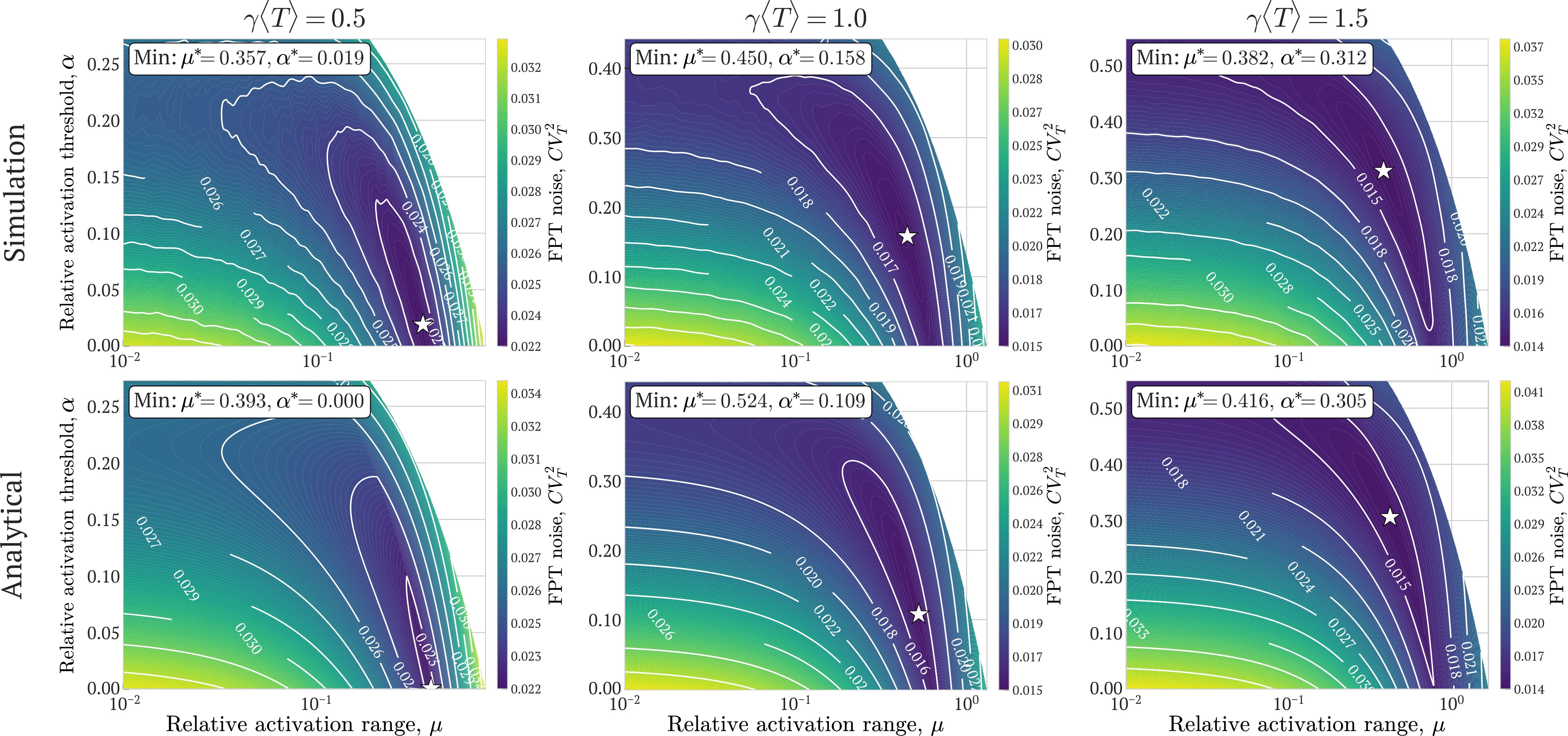}\caption{\small
   \textbf{The FPT noise landscape over $(\alpha, \mu)$ reveals a nontrivial optimal strategy.} FPT noise from simulations (top) and from the analytical expressions (bottom) for different values of $\gamma\langle T\rangle$. For each pair $(\alpha, \mu)$, the downstream threshold $Y$ is adjusted to maintain the target mean FPT $\langle T\rangle$. For each $\gamma\langle T\rangle$, $\alpha$ ranges from $0$ to $0.7\,\alpha_{\max}$, and $\mu$ ranges from values at which the system remains in partial activation to values that give a nearly switch-like response. Simulations use $10^6$ Monte Carlo samples, and analytical results use Eqs.~\eqref{eq:cv2_unsat} and~\eqref{eq:cv2_sat}. The star marks the minimum noise in each panel. For $\gamma\langle T\rangle \geq 1.0$, the minimum lies in the interior of the parameter space, showing that neither limiting case is optimal for identical genes. \textbf{Parameters:} $\gamma = 0.05\,\text{min}^{-1}$, $k_y=10\,\text{min}^{-1}$, $\langle b_x\rangle = \langle b_y\rangle = 5$, $\langle b_x^2\rangle = \langle b_y^2\rangle = 50$, $\bar{x} = 1000$.
}
    \label{fig:heatmap}
\end{figure*}

\section{Optimal strategy as a function of the relative mean FPT}\label{global_optimum}

Optimizing over both activation parameters, we examine how the optimal strategy $(\alpha^*, \mu^*)$ changes with the relative mean FPT $\gamma\langle T\rangle$ (Fig.~\ref{fig:heatmap}). For each $\gamma\langle T\rangle$, the analytical expressions reproduce the simulated noise landscape, including the location of the minimum and the shape of the contours, and they capture how this minimum moves as $\gamma\langle T\rangle$ increases. As in Fig.~\ref{fig:analytical_simulations}, their accuracy improves at higher $\alpha$.

The optimal threshold $\alpha^*$ increases monotonically with $\gamma\langle T\rangle$: the later the event, the later the downstream gene should be activated. This trend agrees with the switch-like limit [Eq.~\eqref{eq:alpha_opt_switch}], in which the optimal threshold also grows with the mean FPT.

The optimal activation range $\mu^*$, in contrast, varies non-monotonically and is largest at $\gamma\langle T\rangle = 1.0$ among the three values we simulated. This behavior results from two opposing effects of timing the event later. First, a later event leaves more time for the activation to rise. At a fixed threshold, the widest activation range that still saturates before the event, set by the regime boundary [Eq.~\eqref{eq:regime_boundary}], therefore increases with $\gamma\langle T\rangle$. Second, a later event raises the optimal threshold $\alpha^*$, and a higher threshold favors narrower ranges, both because it reduces the widest range that still saturates and because it increases the upstream contribution to the noise (Sec.~\ref{sec:opt_mu}). The first effect dominates at small $\gamma\langle T\rangle$. At $\gamma\langle T\rangle = 0.5$, the optimal strategy lies close to the linear-saturated result [Eq.~\eqref{eq:mu_opt_linear}], and up to $\gamma\langle T\rangle = 1.0$, $\mu^*$ stays at or near the regime boundary, so it widens together with this boundary. The second effect dominates at larger $\gamma\langle T\rangle$. At $\gamma\langle T\rangle = 1.5$, the optimal threshold is high enough that $\mu^*$ moves away from the boundary, well inside the full activation regime, and the optimal range narrows.

\section{Conclusions}

In this work, we derived closed-form approximations for the FPT noise of a two-gene timing system with graded activation [Eqs.~\eqref{eq:cv2_unsat} and~\eqref{eq:cv2_sat}]. The key step is to replace the sigmoidal Hill function with a piecewise-linear activation function, which keeps the moment equations closed. The mean and variance of the downstream protein then follow analytically (Sec.~\ref{sec:fpt_noise}), and the small-noise approximation~\eqref{relating_protein_FPT}~\cite{co2017stochastic} converts them into the FPT noise. The resulting expressions hold for any activation threshold and slope, include the dilution of both proteins, and separate the contributions of upstream and downstream fluctuations. They agree with stochastic simulations and capture where the noise is minimal and how this minimum moves with the relative mean FPT (Figs.~\ref{fig:analytical_simulations} and~\ref{fig:heatmap}).

For two identical genes, these expressions lead to three main conclusions about how the shape of the activation function sets timing precision (Secs.~\ref{sec:opt_mu} and~\ref{global_optimum}). First, timing is most precise when the downstream gene reaches full activation no later than the event. If the event occurs while the activation is still rising, widening the activation range only increases the noise. Second, the activation threshold and range must be tuned together, because widening the activation range has two opposing effects. It reduces the downstream contribution to the noise, but it also lets upstream fluctuations modulate the production of the downstream gene for longer, which increases the upstream contribution. Since the upstream contribution is larger the later the activation starts, low thresholds are best paired with graded responses and high thresholds with switch-like ones. Third, a graded response outperforms switch-like activation at every relative mean FPT we explored, and the way its optimal shape changes with the relative mean FPT follows from the first two. A later event leaves more time for the activation to rise, so wider ranges can still reach full activation in time (first conclusion), but it also raises the optimal threshold, which favors narrower ranges (second conclusion). The competition between these two effects makes the optimal activation range first widen and then narrow as the relative mean FPT increases.

These results are consistent with, and complement, those of Gupta \textit{et al.}~\cite{gupta2018temporal,gupta2020temporal}. In the absence of feedback, they found that the optimal regulation function increases gradually with the regulator~\cite{gupta2020temporal}, in line with our finding that a graded response outperforms the switch-like limit. Our expressions show this analytically for a diluted downstream protein and also identify the combination of threshold and range that minimizes the noise. In the switch-like limit, they reduce to the results of our previous work~\cite{hernandez2025cascade}.

Our results rely on two approximations. First, the small-noise approximation treats each threshold crossing as a small deviation from the mean trajectory. It therefore loses accuracy in two cases: when $Y$ is close to the steady state $\bar{y}$, as at low activation thresholds and narrow activation ranges (Fig.~\ref{fig:analytical_simulations}), and when $Y$ is so low that it is reached after only a few bursts. Second, the moments are obtained by applying to every cell the piece of the activation function reached by the mean upstream concentration. This neglects cells whose upstream concentration lies on a different piece near the activation and saturation times. The simulations make neither approximation, so their agreement with the analytical expressions shows that these effects remain small around the optimal strategies.

The framework can be extended in two directions. First, Eqs.~\eqref{eq:cv2_unsat} and~\eqref{eq:cv2_sat} already hold for genes with different intrinsic noises. Under switch-like activation, the ratio $\mathrm{CV}_y/\mathrm{CV}_x$, together with the relative mean FPT, determines whether the optimal timing system uses one gene or both~\cite{hernandez2025cascade}. Under graded activation, the trade-off described above suggests that a noisier upstream gene would favor sharper activation and a noisier downstream gene a more graded one, so the intrinsic noise ratio would also set the shape of the optimal activation function. Determining these optimal strategies, together with the conditions under which coupling two genes improves timing precision, is the subject of ongoing work. Second, the approach is not restricted to two genes. In a longer activation cascade with piecewise-linear activation, the production rate of each gene is still either constant or linear in the concentration of the preceding gene. The moment equations can therefore be solved stage by stage, as in the two-gene case. A longer cascade only adds more regimes, which makes the algebra more involved but does not change the approach.

Synthetic timing systems offer a direct test of these predictions. Pinto \textit{et al.}~\cite{pinto2018engineering} engineered timing circuits with a tunable delay by chaining up to three orthogonal $\sigma$ factors in \textit{E.~coli} and \textit{B.~subtilis}. These $\sigma$ factors activate their targets without cooperativity, so each stage has a graded response like the one studied here, and the extension to longer cascades described above applies directly to them. Measuring the cell-to-cell variability of their response times, and not only their mean delay, would provide this test. More broadly, a piecewise-linear activation function keeps the essential feature of a graded response while still giving closed-form results. This makes it a useful starting point for studying how the shape of gene activation sets the precision of cellular timing in more complex systems.

\newpage

\appendix
\counterwithin*{equation}{section}
\renewcommand{\theequation}{\thesection\arabic{equation}}
\section*{Appendix}
\section{Derivation of the downstream moments}\label{app:moments}

Here, we derive the mean and variance of the downstream protein $y$ reported in the main text [Eqs.~\eqref{eq:mean_y_lin}, \eqref{eq:var_y_lin}, \eqref{eq:mean_y_sat}, and~\eqref{eq:var_y_sat}], starting from the moment equations~\eqref{eq:dynkin}. The upstream equation~\eqref{eq:dynkin_x} is closed on its own and yields Eq.~\eqref{eq:x_moments}. We assign to every cell the piece of $A(x)$ reached by the mean upstream concentration $\langle x\rangle$. Because the activation function~\eqref{activacion_piecewise} is linear within each piece, the averages on the right-hand sides of Eqs.~\eqref{eq:dynkin_y} and~\eqref{eq:dynkin_xy} then reduce to known moments of $x$ and lower-order moments of $y$. The system therefore closes exactly in each regime. The only difference between the two regimes is which linear piece of $A(x)$ applies when the downstream threshold is crossed.

\subsection{Partial activation regime}

When the crossing occurs before saturation, the downstream gene is driven by the linear piece of the activation function~\eqref{activacion_piecewise}, valid for $t_a<t<t_s$. Measuring time from activation through $\tau$, as in Eq.~\eqref{eq:mean_y_lin}, and using $e^{-\gamma t_a}=1-\alpha$ from Eq.~\eqref{eq:times}, the coupling term becomes
\begin{equation}\label{eq:app_coupling}
\langle x\rangle-K=\bar{x}(1-\alpha)\left(1-e^{-\gamma\tau}\right).
\end{equation}
For $n=1$, Eq.~\eqref{eq:dynkin_y} gives the equation for the mean,
\begin{equation}\label{eq:app_mean_ode}
\frac{d\langle y\rangle}{dt}=\lambda\langle b_y\rangle\big(\langle x\rangle-K\big)-\gamma\langle y\rangle,
\end{equation}
a linear first-order differential equation with initial condition $\langle y(t_a)\rangle=0$. Using Eq.~\eqref{eq:app_coupling} and the integrating factor $e^{\gamma\tau}$, we obtain
\begin{equation}
\langle y\rangle=e^{-\gamma\tau}\!\int_0^\tau\! e^{\gamma\tau'}\,\lambda\langle b_y\rangle(1-\alpha)\bar{x}\big(1-e^{-\gamma\tau'}\big)\,d\tau',
\end{equation}
which yields Eq.~\eqref{eq:mean_y_lin}. The variance requires the second moment $\langle y^2\rangle$. Through Eq.~\eqref{eq:dynkin_y} with $n=2$, $\langle y^2\rangle$ couples to the mixed moment $\langle xy\rangle$, whose equation follows from Eq.~\eqref{eq:dynkin_xy} with $m=n=1$
\begin{subequations}\label{eq:app_partial_chain}
\begin{align}
\frac{d\langle xy\rangle}{dt}=&\;k_x\langle b_x\rangle\langle y\rangle+\lambda\langle b_y\rangle\big(\langle x^2\rangle-K\langle x\rangle\big)-2\gamma\langle xy\rangle,\\
\frac{d\langle y^2\rangle}{dt}=&\;2\lambda\langle b_y\rangle\big(\langle xy\rangle-K\langle y\rangle\big)+\lambda\langle b_y^2\rangle\big(\langle x\rangle-K\big)\nonumber\\
&-2\gamma\langle y^2\rangle.
\end{align}
\end{subequations}
The mixed-moment equation depends only on the upstream moments [Eq.~\eqref{eq:x_moments}] and on the mean $\langle y\rangle$ just obtained, so it can be integrated directly using the integrating factor $e^{2\gamma\tau}$ and the initial condition $\langle x(t_a)y(t_a)\rangle=0$. Once $\langle xy\rangle$ is known, the right-hand side of the second-moment equation is fully determined, and this equation is integrated in the same way. Solving the chain $\langle x\rangle\to\langle y\rangle\to\langle xy\rangle\to\langle y^2\rangle$ sequentially and subtracting $\langle y\rangle^2$ from $\langle y^2\rangle$ gives the variance~\eqref{eq:var_y_lin}.

\subsection{Full activation regime}

When the crossing occurs after saturation, the downstream gene is on the constant piece of the activation function~\eqref{activacion_piecewise} for $t>t_s$. Because the rate is then constant, Eq.~\eqref{eq:dynkin_y} no longer couples to $x$, and the moments of $y$ obey the equations of a constitutive gene, starting from their values at the saturation time $t_s$. The mean satisfies
\begin{equation}\label{eq:app_mean_ode_sat}
\frac{d\langle y\rangle}{dt}=k_y\langle b_y\rangle-\gamma\langle y\rangle,
\end{equation}
now with the nonzero initial value $\langle y(t_s)\rangle$ reached at the end of the linear activation phase. We obtain this value by evaluating Eq.~\eqref{eq:mean_y_lin} at $\tau=t_s-t_a$, where $\gamma(t_s-t_a)=\ln[(1-\alpha)/(1-\alpha-\mu)]$ from Eq.~\eqref{eq:times}. Using Eqs.~\eqref{eq:steady_states} and~\eqref{eq:dimless} to write $\lambda\langle b_y\rangle\bar{x}/\gamma=\bar{y}/\mu$, this gives
\begin{equation}\label{eq:app_yts}
\langle y(t_s)\rangle=\bar{y}\left(1-\frac{\bar f}{\mu}\right),
\end{equation}
with $\bar f$ given by Eq.~\eqref{eq:fbar}. Starting from this value, the solution of Eq.~\eqref{eq:app_mean_ode_sat} relaxes toward the saturated steady state $\bar{y}$
\begin{equation}\label{eq:app_relax}
\langle y\rangle=\bar{y}+\big(\langle y(t_s)\rangle-\bar{y}\big)e^{-\gamma(t-t_s)},
\end{equation}
which, combined with Eq.~\eqref{eq:app_yts}, gives Eq.~\eqref{eq:mean_y_sat}. The variance obeys the constitutive equation with constant rate $k_y$,
\begin{equation}\label{eq:app_var_ode_sat}
\frac{d\sigma^2_y}{dt}=k_y\langle b_y^2\rangle-2\gamma\,\sigma^2_y,
\end{equation}
whose solution with initial value $\sigma^2_y(t_s)$ is Eq.~\eqref{eq:var_y_sat}. The first term of Eq.~\eqref{eq:var_y_sat} is the variance that the gene builds up under constitutive production. The second term is the variance accumulated during the graded rise, which decays as $y$ approaches its saturated steady state.

In both regimes, the FPT noise follows from the small-noise approximation~\eqref{relating_protein_FPT}. Substituting the mean and variance of the corresponding regime and evaluating them at the crossing time $t=\langle T\rangle$, estimated from Eq.~\eqref{eq:mean_fpt}, gives Eqs.~\eqref{eq:cv2_unsat} and~\eqref{eq:cv2_sat} for the partial and full activation regimes, respectively. The two expressions coincide at the regime boundary, where the crossing occurs exactly at saturation (Sec.~\ref{sec:opt_mu}).

\bibliographystyle{unsrt}
\bibliography{References}

@article{singh2014stochastic,
  title   = {Stochastic holin expression can account for lysis time variation in the bacteriophage $\lambda$},
  author  = {Singh, Abhyudai and Dennehy, John J.},
  journal = {Journal of the Royal Society Interface},
  volume  = {11},
  number  = {95},
  pages   = {20140140},
  year    = {2014}
}

@article{mondal2024molecular,
  title   = {Molecular mechanisms of precise timing in cell lysis},
  author  = {Mondal, Anupam and Teimouri, Hamid and Kolomeisky, Anatoly B.},
  journal = {Biophysical Journal},
  volume  = {123},
  number  = {18},
  pages   = {3090--3099},
  year    = {2024},
  doi     = {10.1016/j.bpj.2024.07.008}
}

@article{kannoly2025translation,
  title   = {Translation efficiency impacts phage lysis timing and its precision in single cells},
  author  = {Kannoly, Sherin and Singh, Kevin and Juman, Naseerah and Islam, Zubaida Marufee and Caballero Quiroga, I{\~n}igo and Singh, Abhyudai and Dennehy, John J.},
  journal = {bioRxiv},
  year    = {2025},
  doi     = {10.1101/2025.05.14.654169}
}

@article{vandal2026role,
  title   = {The role of cell growth rate on accumulation of the mitotic cyclin {Cdc13} in fission yeast},
  author  = {Vandal, Sarah E. and Rezaee, Sayeh and Nieto, C{\'e}sar and Flynn, Mackenzie J. and Singh, Abhyudai and Moseley, James B.},
  journal = {bioRxiv},
  year    = {2026},
  doi     = {10.64898/2026.05.14.724355}
}

@article{patterson2019noisy,
  title   = {Noisy cell-size-correlated expression of cyclin {B} drives probabilistic cell-size homeostasis in fission yeast},
  author  = {Patterson, James O. and Rees, Paul and Nurse, Paul},
  journal = {Current Biology},
  volume  = {29},
  number  = {8},
  pages   = {1379--1386},
  year    = {2019},
  doi     = {10.1016/j.cub.2019.03.011}
}

@article{nachman2007dissecting,
  title   = {Dissecting timing variability in yeast meiosis},
  author  = {Nachman, Iftach and Regev, Aviv and Ramanathan, Sharad},
  journal = {Cell},
  volume  = {131},
  number  = {3},
  pages   = {544--556},
  year    = {2007},
  doi     = {10.1016/j.cell.2007.09.044}
}

@article{kannoly2022optimal,
  title   = {An optimal lysis time maximizes bacteriophage fitness in quasi-continuous culture},
  author  = {Kannoly, Sherin and Singh, Abhyudai and Dennehy, John J.},
  journal = {mBio},
  volume  = {13},
  number  = {3},
  pages   = {e03593-21},
  year    = {2022}
}

@article{Thattai_vanOudenaarden_2001,
  title   = {Intrinsic noise in gene regulatory networks},
  author  = {Thattai, Mukund and van Oudenaarden, Alexander},
  journal = {Proceedings of the National Academy of Sciences},
  volume  = {98},
  number  = {15},
  pages   = {8614--8619},
  year    = {2001},
  doi     = {10.1073/pnas.151588598}
}

@article{Elowitz_Levine_Siggia_Swain_2002,
  title   = {Stochastic gene expression in a single cell},
  author  = {Elowitz, Michael B. and Levine, Arnold J. and Siggia, Eric D. and Swain, Peter S.},
  journal = {Science},
  volume  = {297},
  number  = {5584},
  pages   = {1183--1186},
  year    = {2002},
  doi     = {10.1126/science.1070919}
}

@article{Ozbudak_Thattai_Kurtser_Grossman_vanOudenaarden_2002,
  title   = {Regulation of noise in the expression of a single gene},
  author  = {Ozbudak, Ertugrul M. and Thattai, Mukund and Kurtser, Iren and Grossman, Alan D. and van Oudenaarden, Alexander},
  journal = {Nature Genetics},
  volume  = {31},
  number  = {1},
  pages   = {69--73},
  year    = {2002},
  doi     = {10.1038/ng869}
}

@article{ghusinga2017first,
  title   = {First-passage time approach to controlling noise in the timing of intracellular events},
  author  = {Ghusinga, Khem Raj and Dennehy, John J. and Singh, Abhyudai},
  journal = {Proceedings of the National Academy of Sciences},
  volume  = {114},
  number  = {4},
  pages   = {693--698},
  year    = {2017},
  doi     = {10.1073/pnas.1609012114}
}

@article{co2017stochastic,
  title   = {Stochastic timing in gene expression for simple regulatory strategies},
  author  = {Dal Co, Alma and Cosentino Lagomarsino, Marco and Caselle, Michele and Osella, Matteo},
  journal = {Nucleic Acids Research},
  volume  = {45},
  number  = {3},
  pages   = {1069--1078},
  year    = {2017},
  doi     = {10.1093/nar/gkw1235}
}

@article{kannoly2020optimum,
  title   = {Optimum threshold minimizes noise in timing of intracellular events},
  author  = {Kannoly, Sherin and Gao, Tianhui and Dey, Supravat and Wang, Ing-Nang and Singh, Abhyudai and Dennehy, John J.},
  journal = {iScience},
  volume  = {23},
  number  = {6},
  pages   = {101186},
  year    = {2020},
  doi     = {10.1016/j.isci.2020.101186}
}

@article{rezaee2026event,
  title   = {Event timing precision in stochastic gene expression with bursty production},
  author  = {Rezaee, Sayeh and Bokes, Pavol and Singh, Abhyudai},
  journal = {bioRxiv},
  year    = {2026},
  doi     = {10.64898/2026.10.25.761601}
}

@article{rosenfeld2003response,
  title   = {Response delays and the structure of transcription networks},
  author  = {Rosenfeld, Nitzan and Alon, Uri},
  journal = {Journal of Molecular Biology},
  volume  = {329},
  number  = {4},
  pages   = {645--654},
  year    = {2003}
}

@article{pinto2018engineering,
  title   = {Engineering orthogonal synthetic timer circuits based on extracytoplasmic function $\sigma$ factors},
  author  = {Pinto, Daniela and Vecchione, Stefano and Wu, Hao and Mauri, Marco and Mascher, Thorsten and Fritz, Georg},
  journal = {Nucleic Acids Research},
  volume  = {46},
  number  = {14},
  pages   = {7450--7464},
  year    = {2018},
  doi     = {10.1093/nar/gky614}
}

@article{hernandez2025cascade,
  title   = {Intrinsic limits of timekeeping precision in gene regulatory cascades},
  author  = {Hern{\'a}ndez, Juan Sebasti{\'a}n and Nieto, C{\'e}sar and Pedraza, Juan Manuel and Singh, Abhyudai},
  journal = {arXiv preprint arXiv:2512.20933},
  year    = {2025},
  doi     = {10.48550/arXiv.2512.20933}
}

@article{hooshangi2005ultrasensitivity,
  title   = {Ultrasensitivity and noise propagation in a synthetic transcriptional cascade},
  author  = {Hooshangi, Sara and Thiberge, Stephan and Weiss, Ron},
  journal = {Proceedings of the National Academy of Sciences},
  volume  = {102},
  number  = {10},
  pages   = {3581--3586},
  year    = {2005},
  doi     = {10.1073/pnas.0408507102}
}

@article{Bintu_Buchler_Garcia_Gerland_Hwa_Kondev_Kuhlman_Phillips_2005,
  title   = {Transcriptional regulation by the numbers: applications},
  author  = {Bintu, Lacramioara and Buchler, Nicolas E. and Garcia, Hernan G. and Gerland, Ulrich and Hwa, Terence and Kondev, Jan{\'e} and Kuhlman, Thomas and Phillips, Rob},
  journal = {Current Opinion in Genetics \& Development},
  volume  = {15},
  number  = {2},
  pages   = {125--135},
  year    = {2005},
  doi     = {10.1016/j.gde.2005.02.006}
}

@article{Garcia_Phillips_2011,
  title   = {Quantitative dissection of the simple repression input--output function},
  author  = {Garcia, Hernan G. and Phillips, Rob},
  journal = {Proceedings of the National Academy of Sciences},
  volume  = {108},
  number  = {29},
  pages   = {12173--12178},
  year    = {2011},
  doi     = {10.1073/pnas.1015616108}
}

@article{gupta2018temporal,
  title   = {Temporal precision of regulated gene expression},
  author  = {Gupta, Shivam and Varennes, Julien and Korswagen, Hendrik C. and Mugler, Andrew},
  journal = {PLoS Computational Biology},
  volume  = {14},
  number  = {6},
  pages   = {e1006201},
  year    = {2018},
  doi     = {10.1371/journal.pcbi.1006201}
}

@article{gupta2020temporal,
  title   = {Temporal precision of molecular events with regulation and feedback},
  author  = {Gupta, Shivam and Fancher, Sean and Korswagen, Hendrik C. and Mugler, Andrew},
  journal = {Physical Review E},
  volume  = {101},
  number  = {6},
  pages   = {062420},
  year    = {2020}
}

@article{quee2021ramp,
  title   = {Ramp approximations of sigmoid control functions in gene networks},
  author  = {Quee, G. and Edwards, R.},
  journal = {Physica D: Nonlinear Phenomena},
  volume  = {418},
  pages   = {132840},
  year    = {2021},
  doi     = {10.1016/j.physd.2020.132840}
}

@article{friedman2006linking,
  title   = {Linking stochastic dynamics to population distribution: an analytical framework of gene expression},
  author  = {Friedman, Nir and Cai, Long and Xie, X. Sunney},
  journal = {Physical Review Letters},
  volume  = {97},
  number  = {16},
  pages   = {168302},
  year    = {2006},
  doi     = {10.1103/PhysRevLett.97.168302}
}

@article{singh2010stochastic,
  title   = {Stochastic hybrid systems for studying biochemical processes},
  author  = {Singh, Abhyudai and Hespanha, Jo{\~a}o P.},
  journal = {Philosophical Transactions of the Royal Society A: Mathematical, Physical and Engineering Sciences},
  volume  = {368},
  number  = {1930},
  pages   = {4995--5011},
  year    = {2010},
  doi     = {10.1098/rsta.2010.0211}
}

@inproceedings{soltani2015moment,
  title     = {Moment closure approximations in a genetic negative feedback circuit},
  author    = {Soltani, Mohammad and Vargas-Garc{\'\i}a, C{\'e}sar and Kumar, Niraj and Kulkarni, Rahul and Singh, Abhyudai},
  booktitle = {2015 American Control Conference (ACC)},
  pages     = {4364--4369},
  year      = {2015},
  organization = {IEEE}
}

@article{schnoerr2017approximation,
  title   = {Approximation and inference methods for stochastic biochemical kinetics---a tutorial review},
  author  = {Schnoerr, David and Sanguinetti, Guido and Grima, Ramon},
  journal = {Journal of Physics A: Mathematical and Theoretical},
  volume  = {50},
  number  = {9},
  pages   = {093001},
  year    = {2017}
}

@article{white2011holin,
  title   = {Holin triggering in real time},
  author  = {White, Rebecca and Chiba, Shinobu and Pang, Ting and Dewey, Jill S. and Savva, Christos G. and Holzenburg, Andreas and Pogliano, Kit and Young, Ry},
  journal = {Proceedings of the National Academy of Sciences},
  volume  = {108},
  number  = {2},
  pages   = {798--803},
  year    = {2011},
  doi     = {10.1073/pnas.1011921108}
}

@article{cahill2024spatial,
  title   = {Spatial and temporal control of lysis by the lambda holin},
  author  = {Cahill, Jesse and Holt, Ashley and Theodore, Matthew and Moreland, Russell and O'Leary, Chandler and Martin, Cody and Bettridge, Kelsey and Xiao, Jie and Young, Ry},
  journal = {mBio},
  volume  = {15},
  number  = {2},
  pages   = {e0129023},
  year    = {2024},
  doi     = {10.1128/mbio.01290-23}
}

@article{grima_fpt,
  title   = {A stochastic vs deterministic perspective on the timing of cellular events},
  author  = {Ham, Lucy and Coomer, Megan A. and {\"O}cal, Kaan and Grima, Ramon and Stumpf, Michael P. H.},
  journal = {Nature Communications},
  volume  = {15},
  pages   = {5286},
  year    = {2024},
  doi     = {10.1038/s41467-024-49624-z}
}

@article{kaern2005stochasticity,
  title   = {Stochasticity in gene expression: from theories to phenotypes},
  author  = {K{\ae}rn, Mads and Elston, Timothy C. and Blake, William J. and Collins, James J.},
  journal = {Nature Reviews Genetics},
  volume  = {6},
  number  = {6},
  pages   = {451--464},
  year    = {2005},
  doi     = {10.1038/nrg1615}
}

@incollection{iyerbiswas2016first,
  title     = {First-passage processes in cellular biology},
  author    = {Iyer-Biswas, Srividya and Zilman, Anton},
  booktitle = {Advances in Chemical Physics},
  volume    = {160},
  pages     = {261--306},
  publisher = {Wiley},
  year      = {2016},
  doi       = {10.1002/9781119165156.ch5}
}

@article{amir2007noise,
  title   = {Noise in timing and precision of gene activities in a genetic cascade},
  author  = {Amir, Amnon and Kobiler, Oren and Rokney, Assaf and Oppenheim, Amos B. and Stavans, Joel},
  journal = {Molecular Systems Biology},
  volume  = {3},
  number  = {1},
  pages   = {71},
  year    = {2007},
  doi     = {10.1038/msb4100113}
}

@article{ditalia2007effects,
  title   = {The effects of molecular noise and size control on variability in the budding yeast cell cycle},
  author  = {Di Talia, Stefano and Skotheim, Jan M. and Bean, James M. and Siggia, Eric D. and Cross, Frederick R.},
  journal = {Nature},
  volume  = {448},
  number  = {7156},
  pages   = {947--951},
  year    = {2007},
  doi     = {10.1038/nature06072}
}

@article{rijal2020protein,
  title   = {Protein hourglass: exact first passage time distributions for protein thresholds},
  author  = {Rijal, Krishna and Prasad, Ashok and Das, Dibyendu},
  journal = {Physical Review E},
  volume  = {102},
  number  = {5},
  pages   = {052413},
  year    = {2020},
  doi     = {10.1103/PhysRevE.102.052413}
}

@article{rijal2022exact,
  title   = {Exact distribution of threshold crossing times for protein concentrations: implication for biological timekeeping},
  author  = {Rijal, Krishna and Prasad, Ashok and Singh, Abhyudai and Das, Dibyendu},
  journal = {Physical Review Letters},
  volume  = {128},
  number  = {4},
  pages   = {048101},
  year    = {2022},
  doi     = {10.1103/PhysRevLett.128.048101}
}

@article{Biswas_Dey_Singh_2024,
  title   = {Sequestration of gene products by decoys enhances precision in the timing of intracellular events},
  author  = {Biswas, Kuheli and Dey, Supravat and Singh, Abhyudai},
  journal = {Scientific Reports},
  volume  = {14},
  pages   = {27199},
  year    = {2024},
  doi     = {10.1038/s41598-024-75505-y}
}

@inproceedings{dey2021feedforward,
  title     = {Feedforward genetic circuits regulate the precision of event timing},
  author    = {Dey, Supravat and Kannoly, Sherin and Bokes, Pavol and Dennehy, John J. and Singh, Abhyudai},
  booktitle = {2021 European Control Conference (ECC)},
  pages     = {2127--2132},
  year      = {2021},
  organization = {IEEE},
  doi       = {10.23919/ECC54610.2021.9654985}
}

@article{biggar2001cell,
  title   = {Cell signaling can direct either binary or graded transcriptional responses},
  author  = {Biggar, Stephen R. and Crabtree, Gerald R.},
  journal = {The EMBO Journal},
  volume  = {20},
  number  = {12},
  pages   = {3167--3176},
  year    = {2001},
  doi     = {10.1093/emboj/20.12.3167}
}

@article{pedraza2005noise,
  title   = {Noise propagation in gene networks},
  author  = {Pedraza, Juan M. and van Oudenaarden, Alexander},
  journal = {Science},
  volume  = {307},
  number  = {5717},
  pages   = {1965--1969},
  year    = {2005},
  doi     = {10.1126/science.1109090}
}

@article{Cai_Friedman_Xie_2006,
  title   = {Stochastic protein expression in individual cells at the single molecule level},
  author  = {Cai, Long and Friedman, Nir and Xie, X. Sunney},
  journal = {Nature},
  volume  = {440},
  number  = {7082},
  pages   = {358--362},
  year    = {2006},
  doi     = {10.1038/nature04599}
}

@article{Yu_Xiao_Ren_Lao_Xie_2006,
  title   = {Probing gene expression in live cells, one protein molecule at a time},
  author  = {Yu, Ji and Xiao, Jie and Ren, Xiaojia and Lao, Kaiqin and Xie, X. Sunney},
  journal = {Science},
  volume  = {311},
  number  = {5767},
  pages   = {1600--1603},
  year    = {2006},
  doi     = {10.1126/science.1119623}
}

\end{document}